\documentclass[reprint,aps,prl,twocolumn,superscriptaddress,preprintnumbers,letterpaper,natbib,floatfix,nofootinbib]{revtex4-2}
\usepackage{amsmath,amssymb,amsfonts}
\usepackage[dvips]{color}
\usepackage{xcolor}
\usepackage{dsfont}
\usepackage{mathrsfs}
\usepackage{float}
\usepackage{hyperref}
\usepackage{graphicx}
\usepackage{graphics}
\usepackage{setspace}
\usepackage{lmodern}
\usepackage{tikz}
\usepackage{cancel}
\usepackage{placeins}
\hypersetup{colorlinks,citecolor=blue,urlcolor=blue,linkcolor=blue}

\newcommand\beq{\begin{eqnarray}}
\newcommand\eeq{\end{eqnarray}}

\begin{document}
\title{Precision unification and the scale of supersymmetry}

\author{Prudhvi N.~Bhattiprolu}
\author{James D.~Wells}
\affiliation{Leinweber Center for Theoretical Physics,\\ University of Michigan, Ann Arbor, MI 48109, USA}

\begin{abstract}
In this letter, we study the implications of precise gauge coupling unification on
supersymmetric particle masses. We argue that precise unification favors the
superpartner masses that are in the range of several TeV and well beyond. We
demonstrate this in the minimal supersymmetric theory with a common
sparticle mass threshold, and two simple high-scale scenarios: minimal supergravity and minimal anomaly-mediated supersymmetry. We also identify candidate models with a Higgsino or a wino dark matter candidate. Finally, the analysis shows unambiguously that unless one takes foggy naturalness notions too seriously, the lack of direct superpartner discoveries at the LHC has not diminished the viability of supersymmetric unified theories in general nor even precision unification in particular.
\end{abstract}

\maketitle
%\preprint{LCTP-23-XX} 

{\bf 1.~Introduction.}
Although the data at the Large Hadron Collider (LHC) is consistent with
the Standard Model (SM) and there are no unambiguous signs of new physics
up to the TeV range, supersymmetry still remains a plausible extension to
the SM for several reasons. For example, it can explain the origin of the weak scale \cite{Dimopoulos:1981zb}, provide for the unification of gauge interactions \cite{Dimopoulos:1981yj}, and accommodate a viable dark matter (DM) candidate \cite{Goldberg:1983nd,Ellis:1983ew}.

Both the discovery of SM-like Higgs boson with $m_h \! \sim \! 125$ GeV and
the lack of evidence for superpartners at the LHC (albeit under simplifying assumptions)
suggest that supersymmetry exists above the TeV scale, if it exists at all. It must be kept in mind that all the undiscovered superpartners, unlike the SM fermions and the massive gauge bosons, can get their masses entirely from supersymmetry-breaking terms and therefore in principle can be much heavier than the weak scale. Moreover, supersymmetric theories obey decoupling as the supersymmetry-breaking scale is raised. Some phenomenologically viable scenarios where the gauginos are somewhat above the TeV scale and the sfermions are much heavier at order 100 TeV and beyond can be found, e.g.,
in Refs.~\cite{Wells:2003tf,Arkani-Hamed:2004zhs,Wells:2004di,Arvanitaki:2012ps}.

In this letter, we will explore the implications of the precision unification conjecture: {\it supersymmetry is a correct principle of nature, and the gauge couplings unify at a high scale with high-scale threshold corrections much smaller in magnitude than naive expectations from grand unified theories (GUTs).} 

The reason for this conjecture is that within the minimal supersymmetric theory the three gauge couplings, when renormalization-group flowed to the high scale, meet much more closely than typical threshold corrections of a grand unification group from its remnants of high-dimensional representations~\cite{Ellis:2015jwa,Ellis:2017erg}. The natural implication of this has long been to put more stock in the idea of supersymmetry. But the extraordinary confluence of gauge couplings, which will be demonstrated below, may be giving a message stronger than that: the high-scale threshold corrections are highly suppressed.

One could speculate on reasons for suppressed threshold corrections.  Perhaps the precision unification comes from an orbifold GUT that projects out extraneous representations, which may even arise from string compactification~\cite{Hebecker:2004ce,Raby:2009sf}. We are agnostic about any given precise underlying reason, but we take seriously the hint of precision gauge coupling unification as embodied in the conjecture above.

{\bf 2.~Supersymmetric frameworks.}
Precision unification has been studied in the past. For example, it has been recognized within TeV-scale MSSM that precise unification can follow from small $\mu$-term and disengaging the gluino mass from the normal assumptions of ``GUT-normalized" gaugino mass hierarchies~\cite{Raby:2009sf,Krippendorf:2013dqa}. Such ideas remain valid. However, in the present day we also know that the LHC has not found superpartners and that the Higgs boson mass has a rather high value of $125\,{\rm GeV}$, consistent with heavy sparticle mass spectrum. This allows for a much heavier superpartner spectrum and the prospect of finding precision unification even within the more standard supersymmetric frameworks. The two straightforward approaches to supersymmetry that we employ to investigate the consequences of our precision unification conjecture are minimal supergravity with its GUT-normalized gaugino mass hierarchy (mSUGRA) and minimal anomaly mediation (mAMSB) with its special anomaly-mediated gaugino mass hierarchy. Both of these scenarios are reviewed in detail in~\cite{Martin:1997ns}.

To be more precise, we first study these implications for
the minimal supersymmetric SM (MSSM) with a common superpartner mass threshold.
After that, we consider mSUGRA and mAMSB rigorously. The MSSM particle spectrum can be determined by
just a few parameters, and we use well-tested computational tools in the literature to do multi-loop renormalization group flow and mass determinations. 

Supersymmetry-breaking is gravity-mediated in mSUGRA, and all superpartner masses are somewhat similar in mass.
On the other hand, gaugino masses are mediated via the superconformal anomaly in mAMSB, giving gaugino masses  in a distinctive hierarchy and one-loop order lower than scalar masses~\cite{Giudice:1998xp,Randall:1998uk,Gherghetta:1999sw,Luty:2001zv,Harnik:2002et}.

The Higgs boson mass is an output in the supersymmetric theories, being a function of other masses and couplings already specified by the model. When identifying models with exact gauge unification we require the lightest CP-even neutral Higgs boson mass to be $\sim$ 125 GeV within uncertainties of the calculation.

Furthermore, we also identify regions of parameter space where the lightest neutralino
is the lightest supersymmetric particle (LSP), and can generate the required thermal
abundance of Higgsino or wino DM assuming $R$-parity is conserved
\cite{Wells:2003tf,Wells:2004di,Profumo:2004at,Giudice:2004tc,Pierce:2004mk,Arkani-Hamed:2006wnf,Hisano:2006nn,Cirelli:2007xd,Hryczuk:2010zi,Baer:2011ec,Cohen:2013ama,Fan:2013faa,Baer:2016ucr,Kowalska:2018toh,Co:2021ion,Co:2022jsn}.

%%%%%%%%%%%%%%%%%%%%%%%%%%%%%%%%%%%%%%%%%%%%%%%%%%%%%%%%%%%%%%%%%%%%%%%%%%%%%%%%%%
\begin{figure}[h!]
    \includegraphics[width=0.98\columnwidth]{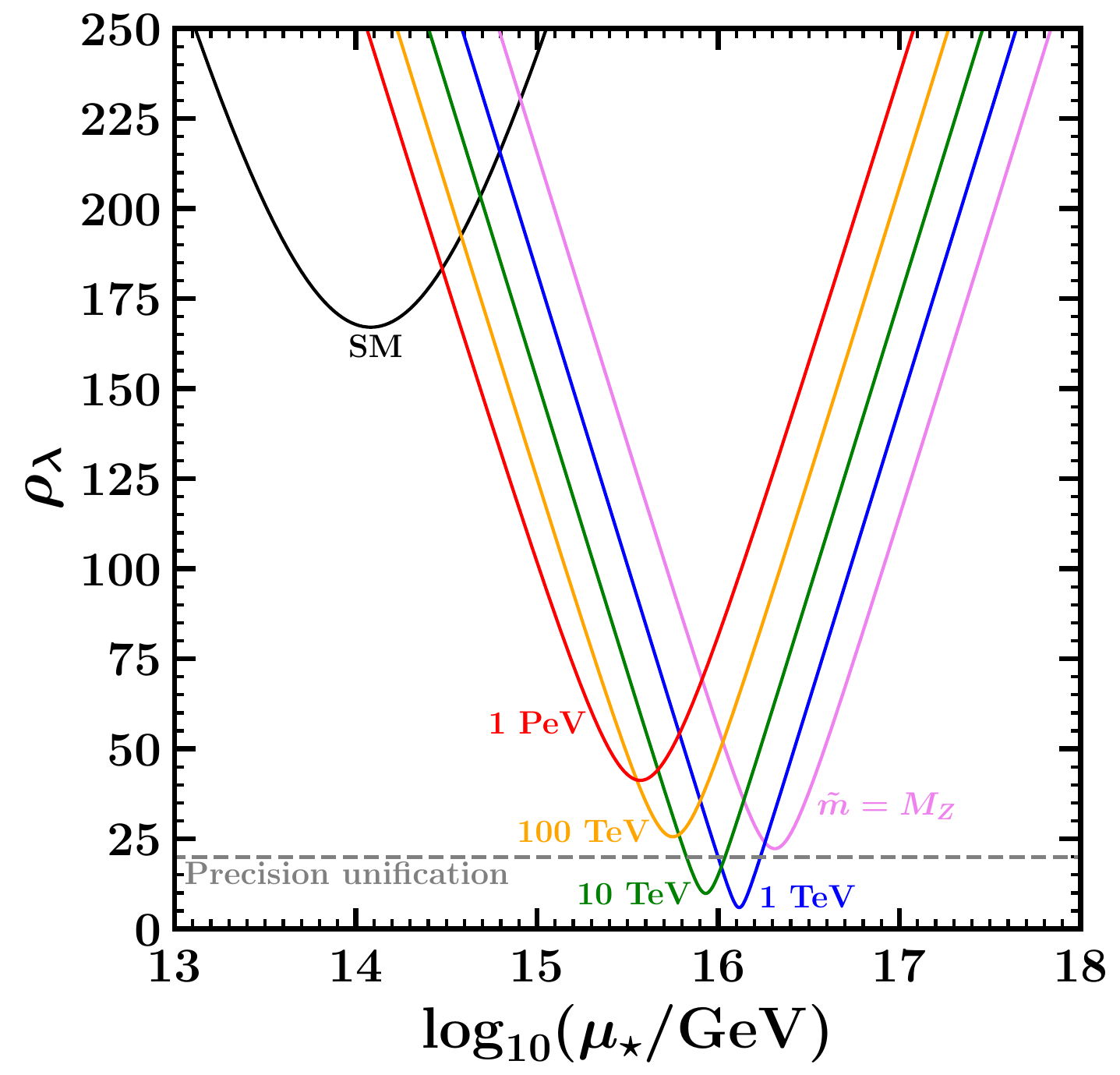}
 \caption{$\rho_\lambda$ as a function of the putative unification scale in the SM (black line)
 and in the MSSM with a common supersymmetric particle mass threshold $\tilde{m}$
 (various colored lines for various $\tilde{m}$ as labeled).
\label{fig:MSSMwithcommonthreshold}}
\end{figure}
%%%%%%%%%%%%%%%%%%%%%%%%%%%%%%%%%%%%%%%%%%%%%%%%%%%%%%%%%%%%%%%%%%%%%%%%%%%%%%%%%% 

{\bf 3.~MSSM with a common threshold.}
It is well-known that, unlike in the SM, the gauge couplings approximately unify in the MSSM
with supersymmetric particle masses roughly around the TeV scale.
As a measure of unification of the gauge couplings, we define
\beq
\frac{\rho_\lambda}{48 \pi^2} &\equiv& \sqrt{\sum_{i \ne j} \left(\frac{1}{g^2_i} - \frac{1}{g^2_j}\right)^2},
\label{eq:rholambda}
\eeq
with $i, j = 1, 2, 3$.
Here, $g_i$ are the gauge couplings with the usual GUT normalization.
The minimum value of $\rho_\lambda$, obtained at the scale $\mu_\star$, is denoted by
$\rho^\text{min}_\lambda$ ($\equiv \rho_\lambda (\mu_\star)$).

Within standard grand unified theories $\rho^\text{min}_\lambda$ is a weighted logarithmic mass sum of remnant high-scale representations.
Specifically, assuming degenerate masses within an irreducible representation,
we have \cite{Hall:1980kf,Ellis:2015jwa}:
\begin{align}
\frac{1}{g^2_i (\mu_\star)} - \frac{1}{g^2_j (\mu_\star)}
&= \left( I^{V_n}_j - I^{V_n}_i\right) \left( 1 + 21 \ln \frac{\mu_\star}{M_{V_n}} \right)
\nonumber \\
&\quad
- \left( I^{S_n}_j - I^{S_n}_i\right) \ln \frac{\mu_\star}{M_{S_n}}
\nonumber \\
&\quad
- 8 \left( I^{F_n}_j - I^{F_n}_i\right) \ln \frac{\mu_\star}{M_{F_n}}
,
\end{align}
with an implicit sum over $n$ different particles for each of the vectors $V_n$, scalars $S_n$, and fermions $F_n$.
Here, $I^{X}_i$ are the Dynkin indices of the representation of $X$ under $(SU(3)_c, SU(2)_L, U(1)_Y)$ for $i = (3, 2, 1)$, respectively.
In typical supersymmetric GUT theories,
such as those discussed in~\cite{Raby:2017ucc},
one expects to have values of $\rho^\text{min}_\lambda$ roughly of order hundreds.

For precision unification, on the other hand, we require $\rho^\text{min}_\lambda < 20$
which roughly corresponds to $3\sigma$ deviation from exact
gauge coupling unification. By that we mean that we allow a factor of three higher correction than what might arise from naive Planck scale corrections:
$\left(\alpha^{-1}_i - \alpha^{-1}_j\right)/4 \pi \sim \mu_\star/M_{P}$
(see also Refs.~\cite{Ellis:2017erg, Ellis:2015jwa}). Such a threshold should not be taken too seriously. There are potential reasons for raising the allowed $\rho^{\rm min}_\lambda$ and for lowering it somewhat to define ``precision unification", but to be concrete we choose $\rho^\text{min}_\lambda < 20$. 

Figure~\ref{fig:MSSMwithcommonthreshold} shows $\rho_\lambda$ as a function of the
putative unification scale in the SM and MSSM with various choices for the
common superpartner threshold $\tilde m$. We performed the renormalization group
evolution (RGE) of the gauge couplings in the (MS)SM at 2-loop (along with 1-loop running of the third generation Yukawa couplings) \cite{Martin:1993zk}.
It is evident from the figure that precision unification is achieved in the MSSM with
a common threshold if $\tilde{m}$ is roughly in the $1 - 10$ TeV range.

%%%%%%%%%%%%%%%%%%%%%%%%%%%%%%%%%%%%%%%%%%%%%%%%%%%%%%%%%%%%%%%%%%%%%%%%%%%%%%%%%%
\begin{figure*}
  \begin{minipage}[]{0.495\linewidth}
    \includegraphics[width=8cm]{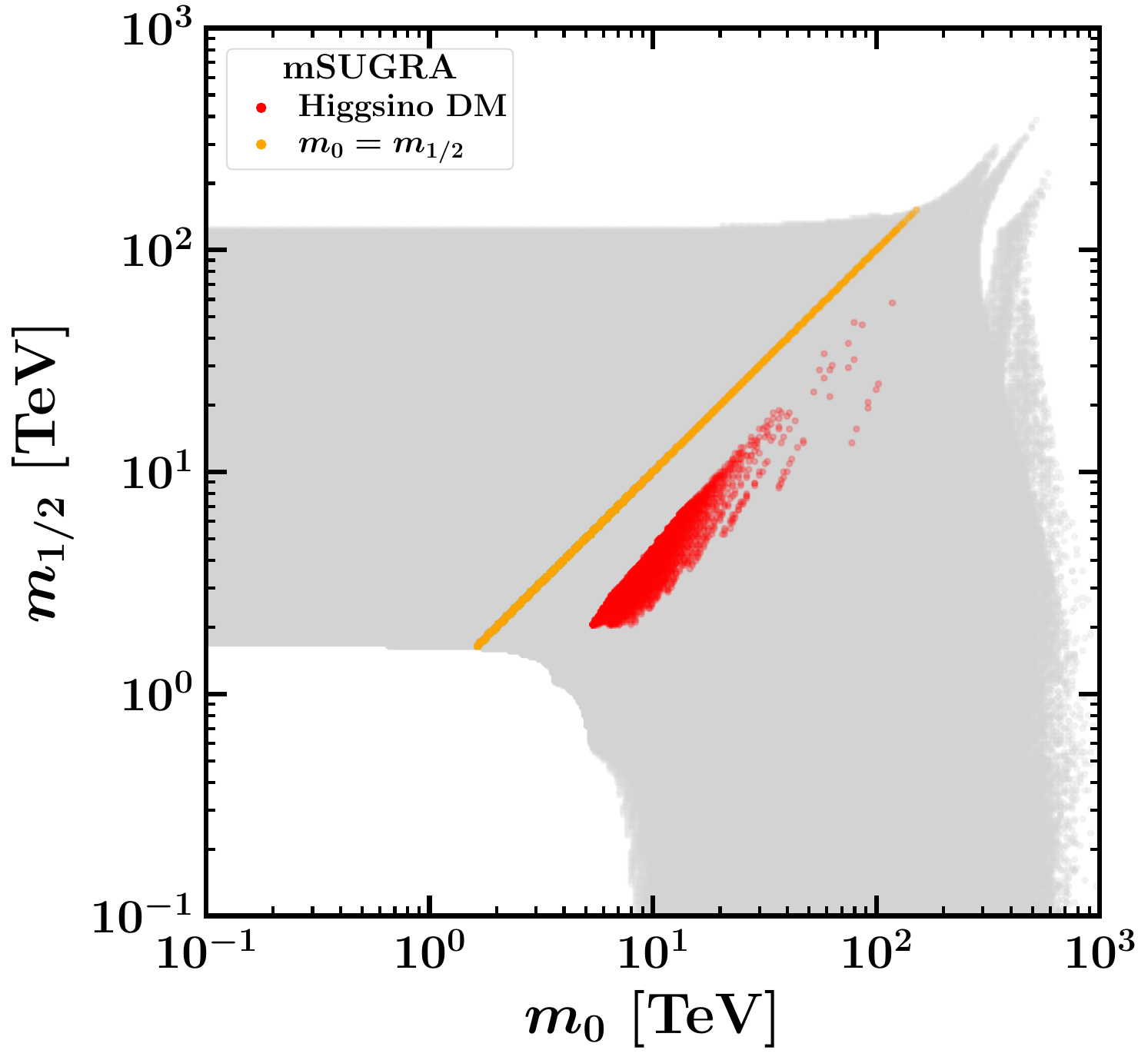}
  \end{minipage}
  \begin{minipage}[]{0.495\linewidth}
    \includegraphics[width=8.6cm]{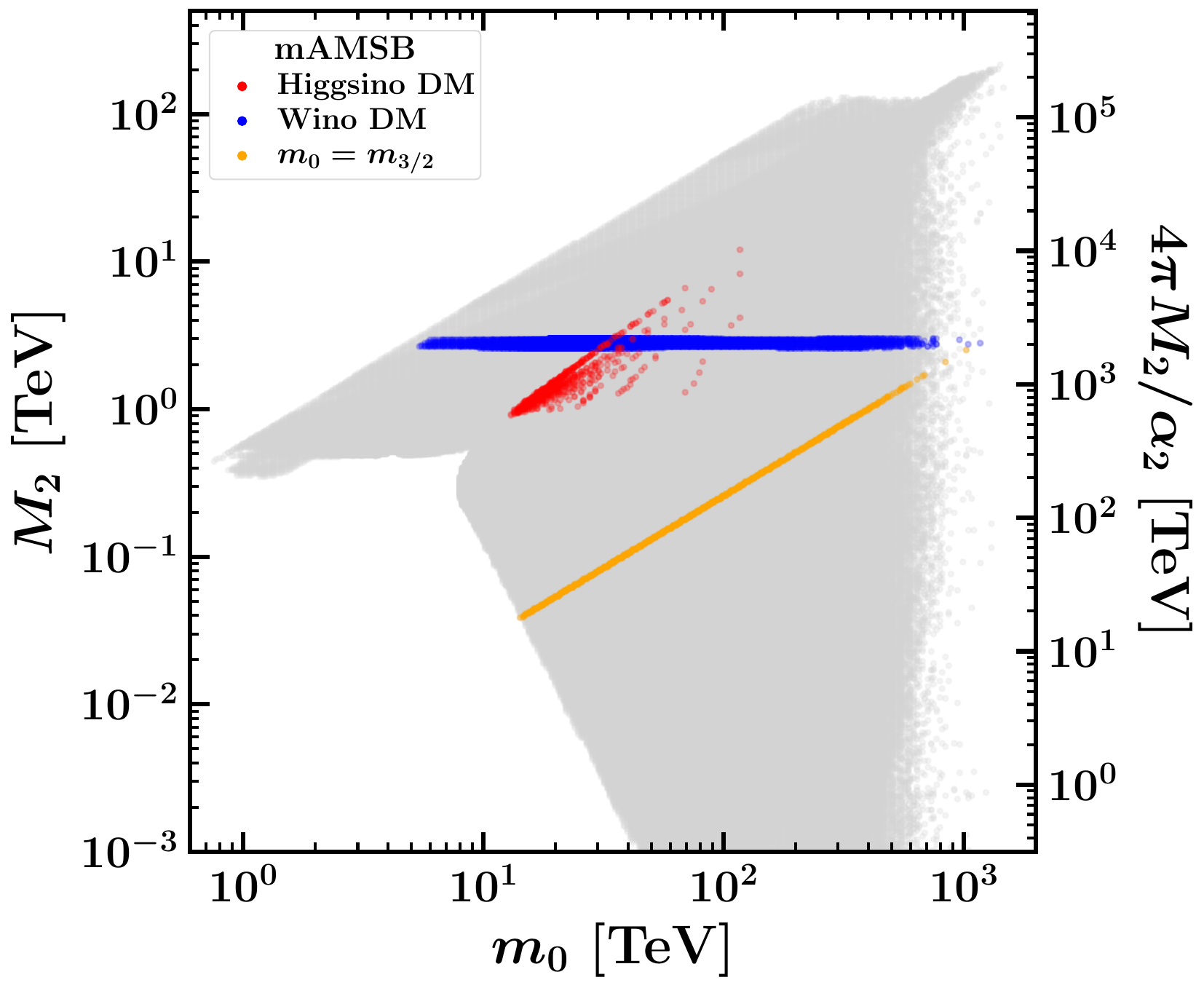}
  \end{minipage}
  \vspace{0.2cm}
  \begin{minipage}[]{0.495\linewidth}
    \includegraphics[width=8cm]{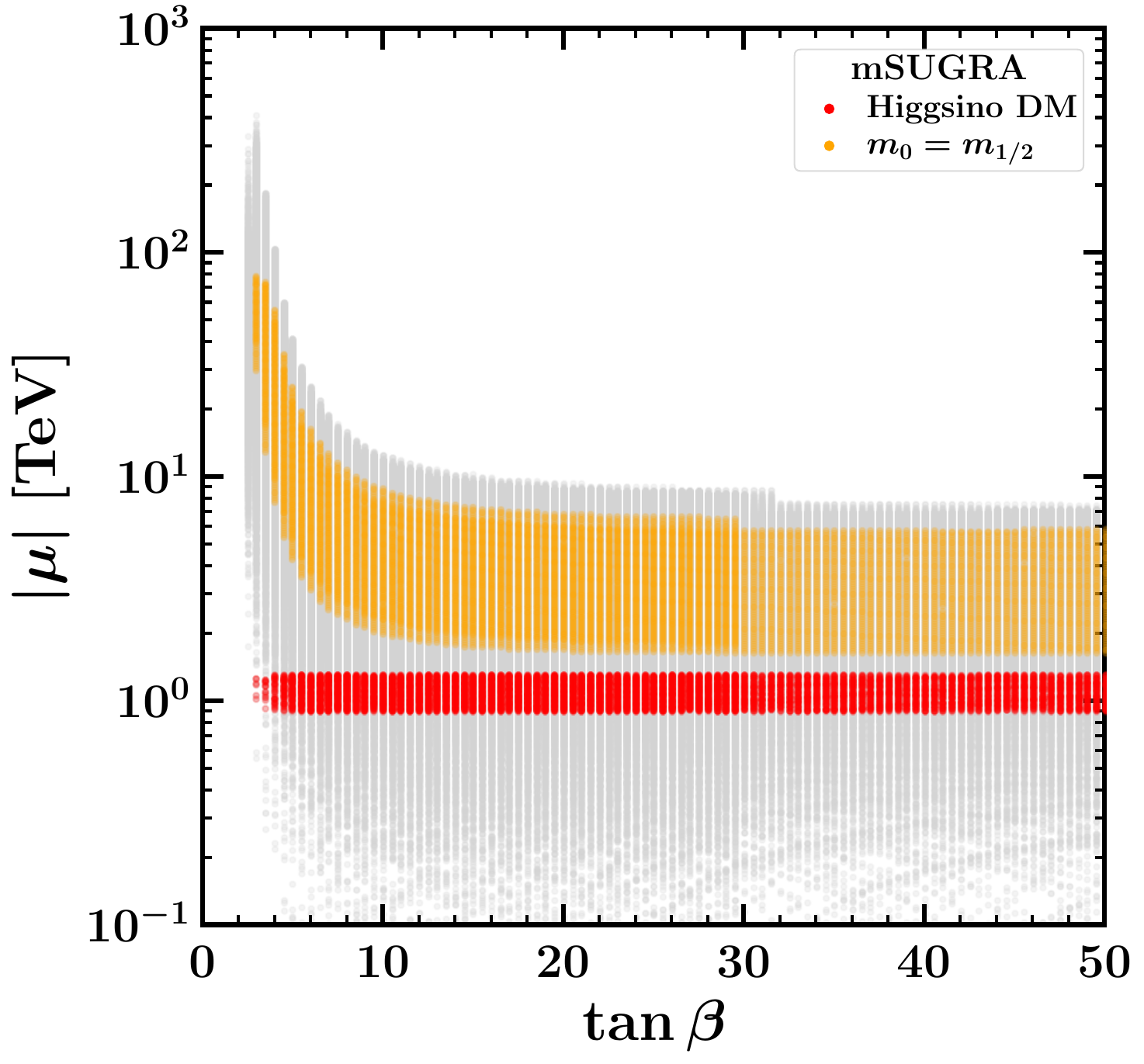}
  \end{minipage}
  \begin{minipage}[]{0.495\linewidth}
    \includegraphics[width=8cm]{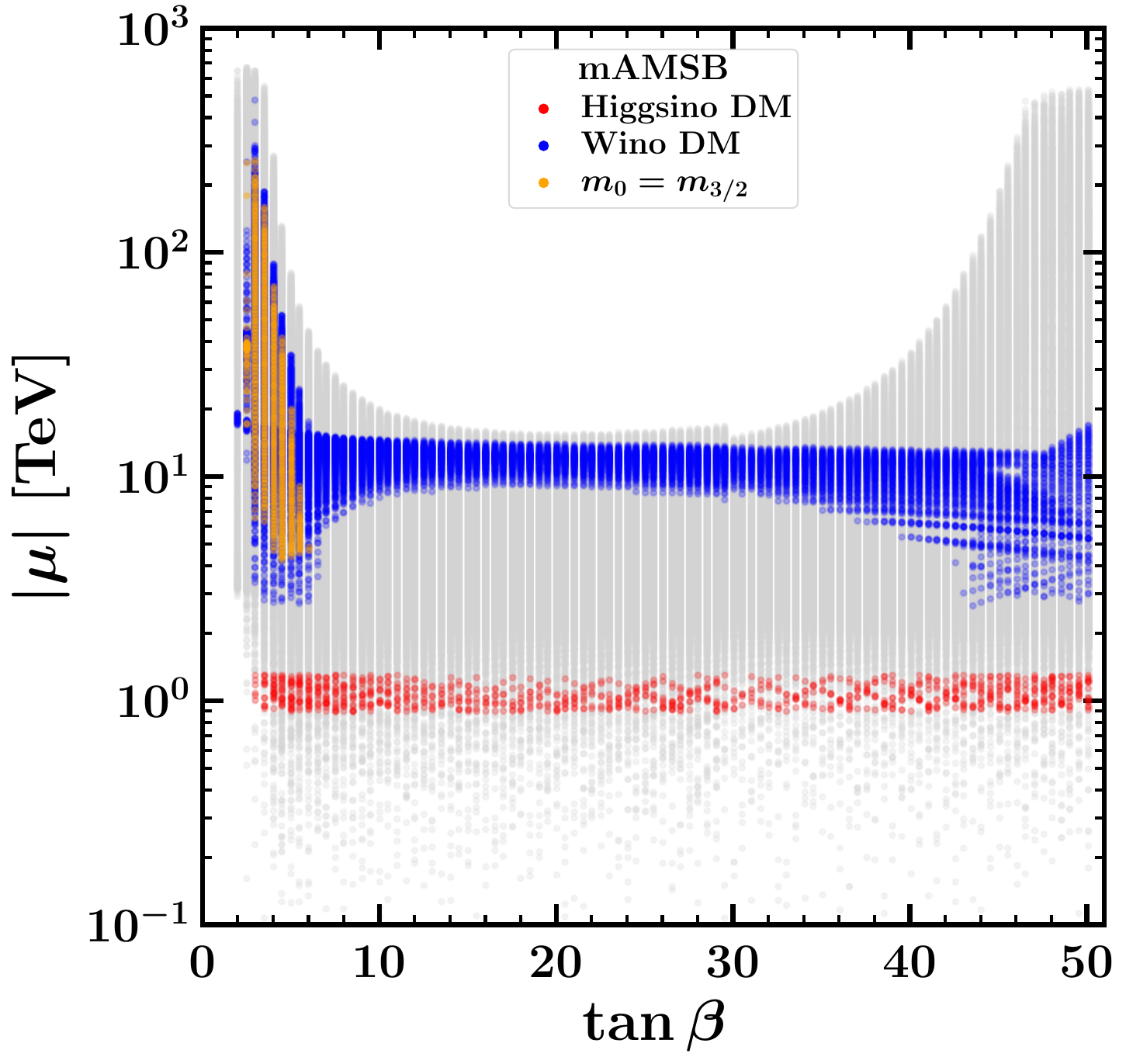}
  \end{minipage}
 \caption{Parameter space (gray points) where precise gauge unification ($\rho_\lambda^\text{min} < 20$)
 can be achieved in addition to satisfying the observed Higgs mass constraint ($m_h = 125.25 \pm 3$ GeV).
 Various colored points correspond to various special cases as labeled.
 Top panels show $m_{1/2}$ in mSUGRA (top-left) and the wino mass $M_2$ in mAMSB (top-right)
 plotted against $m_0$. The right vertical axis in the top-right panel shows a rough estimate of
 $m_{3/2} \sim 4 \pi M_2/\alpha_2$ with $\alpha_2^{-1} \simeq 25$ in mAMSB scenario.
 Bottom panels show the $|\mu|$ term plotted against $\tan \beta$ in mSUGRA (bottom-left) and
 mAMSB (bottom-right) scenarios.
\label{fig:paramspace}}
\end{figure*}
%%%%%%%%%%%%%%%%%%%%%%%%%%%%%%%%%%%%%%%%%%%%%%%%%%%%%%%%%%%%%%%%%%%%%%%%%%%%%%%%%% 

%%%%%%%%%%%%%%%%%%%%%%%%%%%%%%%%%%%%%%%%%%%%%%%%%%%%%%%%%%%%%%%%%%%%%%%%%%%%%%%%%%
\begin{figure*}
  \begin{minipage}[]{0.495\linewidth}
    \centering
    \includegraphics[width=8cm]{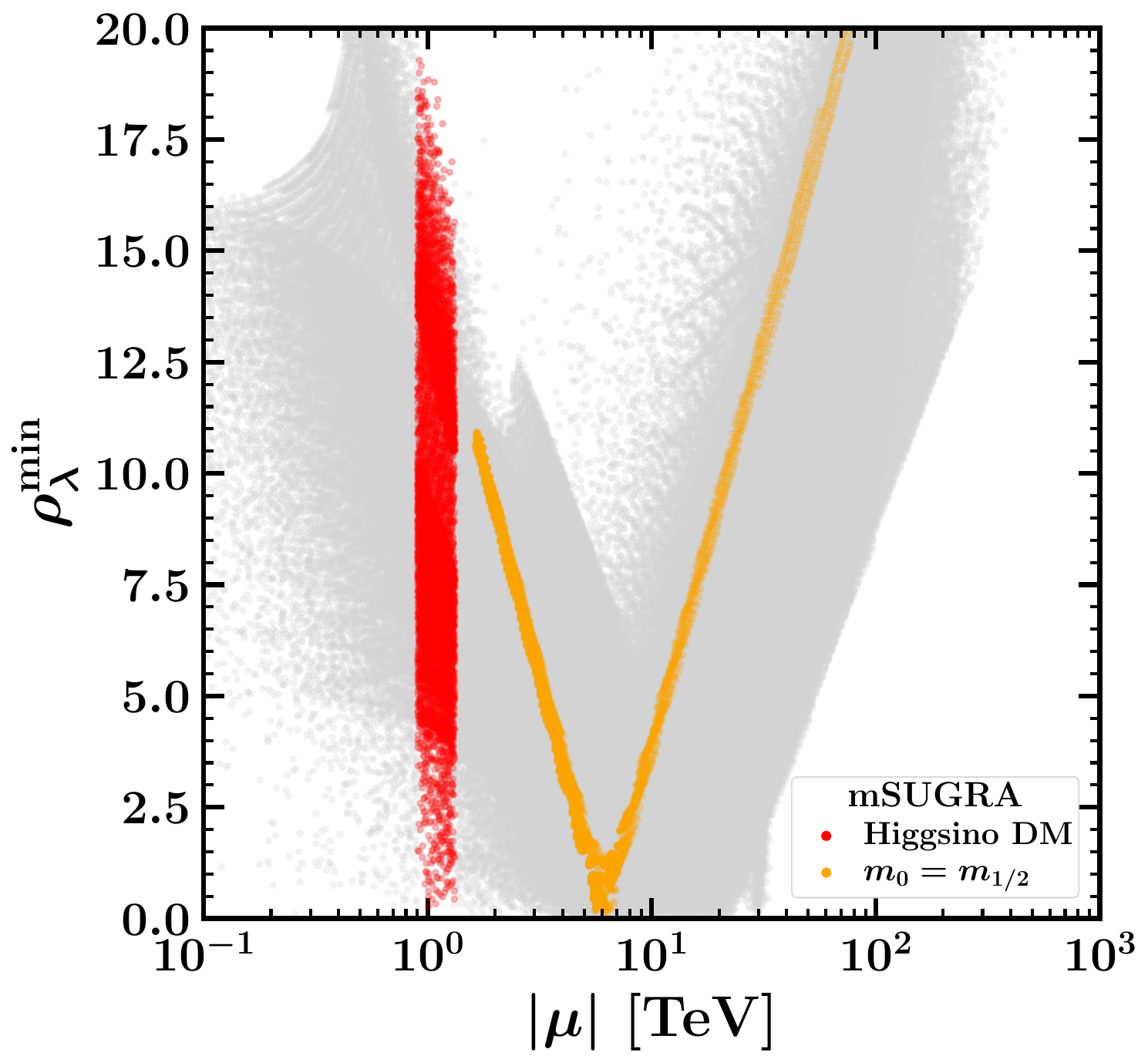}
  \end{minipage}
  \begin{minipage}[]{0.495\linewidth}
    \centering
    \includegraphics[width=8cm]{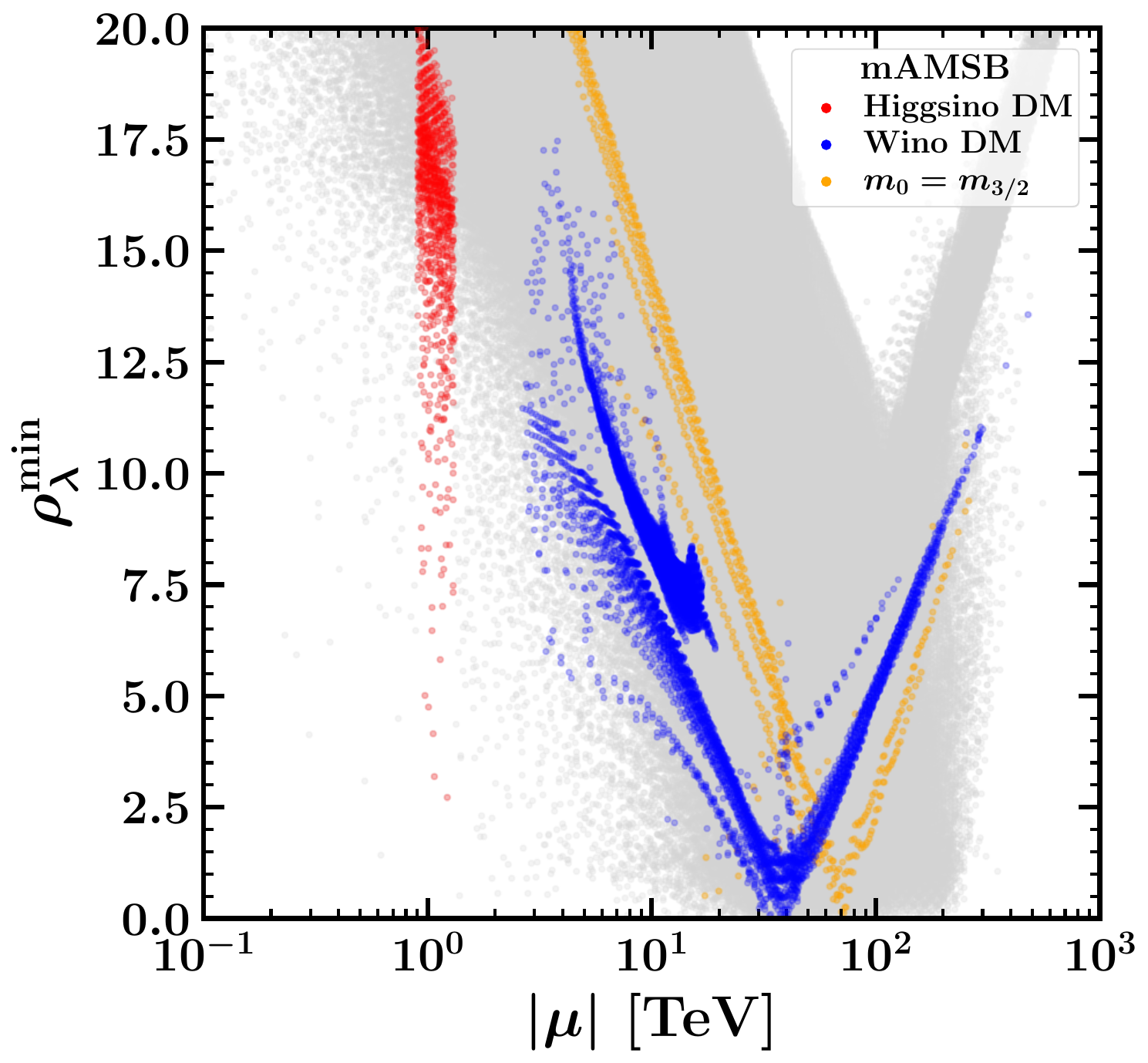}
  \end{minipage}
  \vspace{0.2cm}
  \begin{minipage}[]{0.495\linewidth}
    \centering
    \includegraphics[width=8cm]{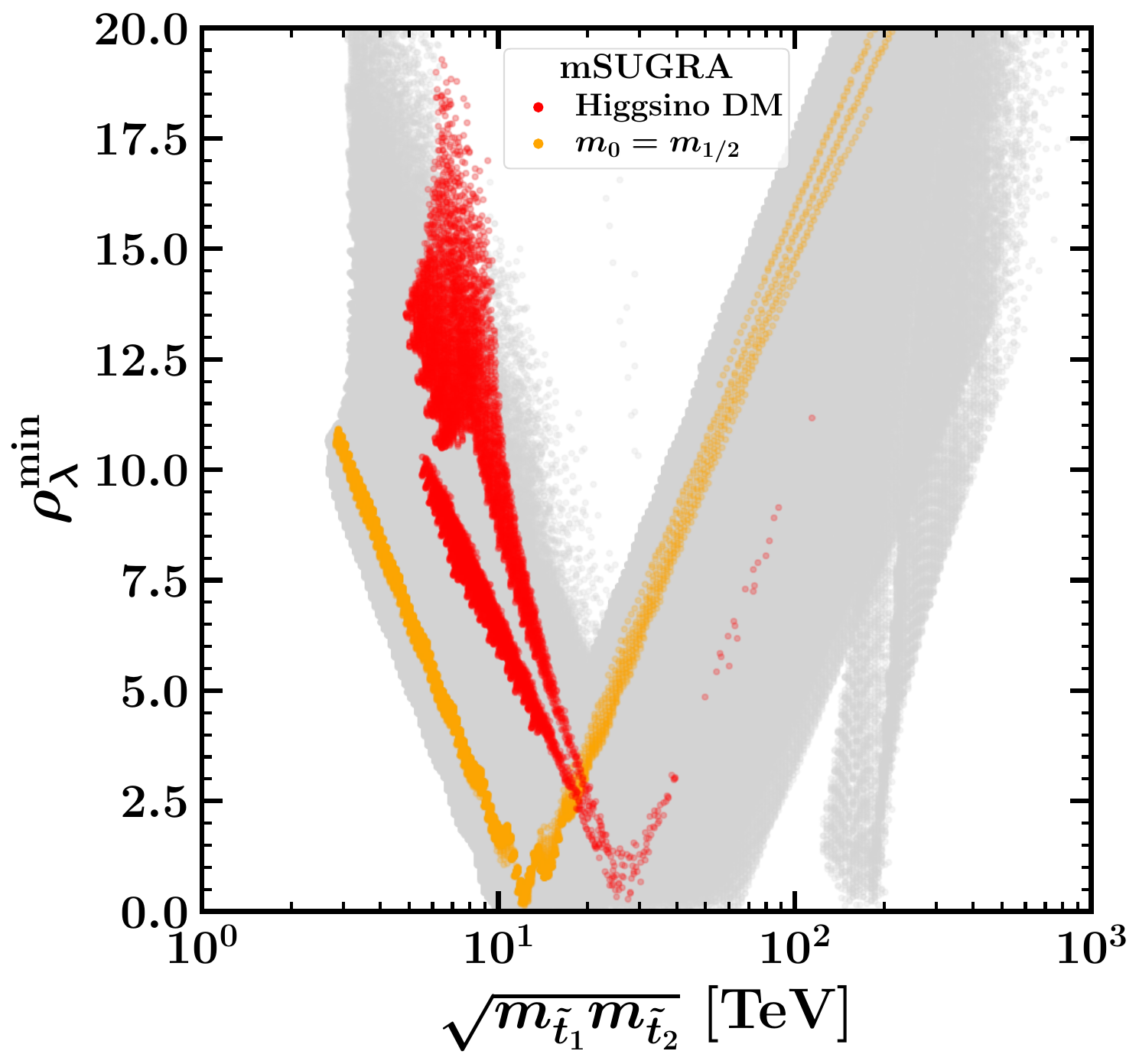}
  \end{minipage}
  \begin{minipage}[]{0.495\linewidth}
    \centering
    \includegraphics[width=8cm]{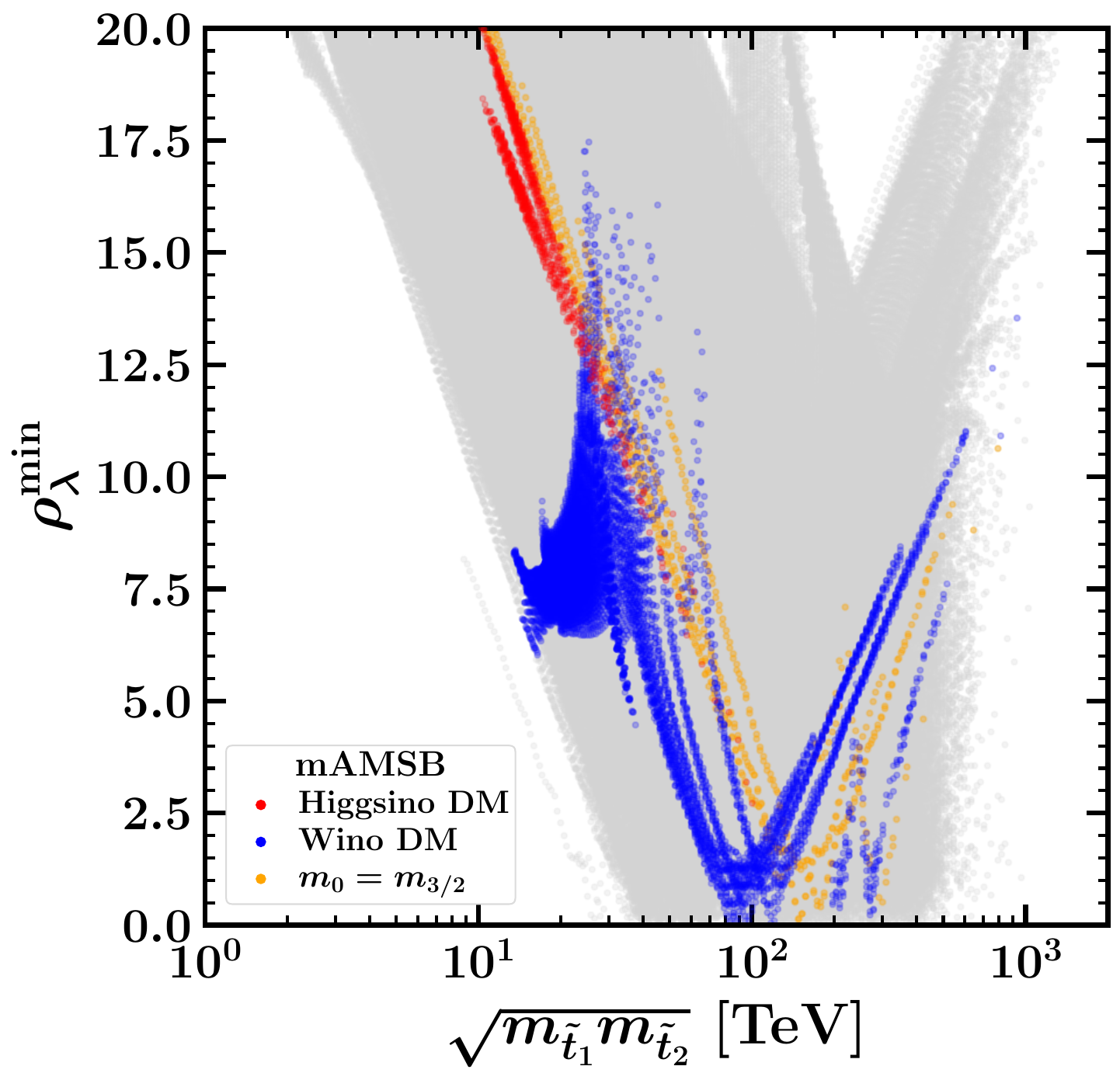}
  \end{minipage}
 \caption{$\rho_{\lambda}^\text{min}$ plotted against the absolute value of the $\mu$ term (top panels) and
 the geometric mean of the top squark masses (bottom panels) in mSUGRA (left) and mAMSB (right) scenarios.
 The gray points correspond to all models with precise gauge coupling unification
 ($\rho_{\lambda}^\text{min} < 20$) and the Higgs mass $m_h$ within 3 GeV of 125.25 GeV.
 Various colored points correspond to various special cases as labeled.
\label{fig:rhomin}}
\end{figure*}
%%%%%%%%%%%%%%%%%%%%%%%%%%%%%%%%%%%%%%%%%%%%%%%%%%%%%%%%%%%%%%%%%%%%%%%%%%%%%%%%%% 

{\bf 4.~High scale scenarios.}
We now turn to the implications of exact gauge unification in mSUGRA and mAMSB frameworks.
Both frameworks have three common parameters: the unified scalar mass $m_0$ at the
GUT scale\footnote{Supersymmetric spectra generators, including {\sc Spheno} \cite{Porod:2011nf,Porod:2003um}, which we have employed, commonly define the GUT scale as the scale where the gauge couplings $g_1$ and $g_2$ unify.},
the ratio of the vacuum expectation values of the two Higgs doublets $\tan \beta$ (at $M_Z$), and
the sign of the Higgsino mass parameter $\mu$.
mSUGRA has two additional parameters, namely, the unified gaugino mass $m_{1/2}$ at the GUT scale and
the universal scalar trilinear coupling $A_0$ at the GUT scale.
On the other hand, mAMSB has one additional parameter: the gravitino mass $m_{3/2}$ at the GUT scale.
In mSUGRA (mAMSB), bino (wino) is the lightest gaugino.
Therefore, the LSP
can be bino-like (wino-like) or Higgsino-like in mSUGRA (mAMSB),
depending on $\mu$ and $\tan \beta$.

We used {\sc Spheno} \cite{Porod:2011nf,Porod:2003um} for generating the MSSM particle spectrum,
which implements 2-loop supersymmetric RGEs (and 3-loop SM RGEs) with improved prediction of
the Higgs boson mass \cite{Staub:2017jnp}. To find the parameter space of interest in these
model frameworks, we independently varied $(m_0, m_{1/2})$ and $(m_0, m_{3/2})$ in mSUGRA and
mAMSB scenarios, respectively, from $10^2$ to $10^8$ GeV in 500 evenly spaced steps on a log scale.
We simultaneously also varied $\tan \beta$ from 2 to 50 in steps of 0.5 for both $\mu \gtrless 0$.
In mSUGRA, we found that varying $A_0$ did not have much impact on our results,
especially when the scale of supersymmetry is much larger than $M_Z$.
Therefore we have simply taken $A_0 = 0$.
In order to impose the observed Higgs boson mass constraint, we required that
$m_h = 125.25 \pm 3$ GeV for the theory calculation. Requiring that the models reproduce the observed Higgs boson mass is
a non-trivial constraint and significantly restricts the parameter space.
For precision unification, we impose $\rho^\text{min}_{\lambda} < 20$.
Moreover, we also identify models where the required thermal abundance is generated by Higgsino or wino DM.
In particular, we identify models with a neutralino LSP and require
$\mu = 1.1 \pm 0.2$ TeV (with $|\mu| < M_1, M_2$)
\cite{Profumo:2004at,Giudice:2004tc,Pierce:2004mk,Hryczuk:2010zi}
for Higgsino DM, and
$M_2 = 2.8 \pm 0.2$ TeV (with $M_2 < M_1, |\mu|$)
\cite{Hryczuk:2010zi,Hisano:2006nn,Cirelli:2007xd}
for wino DM.

In Figure~\ref{fig:paramspace}, we show the parameter space (gray points)
in both mSUGRA (left panels) and mAMSB (right panels) frameworks with precise
gauge coupling unification ($\rho_\lambda^\text{min} < 20$) and the observed
Higgs boson mass within 3 GeV of 125.25 GeV.
Specifically, $m_0$ vs $m_{1/2}$ (top-left panel) and $|\mu|$ vs $\tan \beta$ (bottom-left panel)
scatter plots in mSUGRA. And $m_0$ vs $M_2$ (or equivalently $m_{3/2} \sim 4 \pi M_2/\alpha_2$)
(top-right panel) and $|\mu|$ vs $\tan \beta$ (bottom-right panel) scatter plots in mAMSB.
Figure~\ref{fig:paramspace} also shows the parameter space where the neutralino LSP,
that can reproduce the required thermal DM abundance, is Higgsino-like (red points) or
wino-like (blue points).
In mSUGRA, since the neutralino LSP cannot be wino-like, there are no cases with wino DM.
Also shown in the figure are the cases (orange points) where
$m_0 = m_{1/2}$ in mSUGRA (left panels) and $m_0 = m_{3/2}$ in mAMSB (right panels).

Finally, Figure~\ref{fig:rhomin} shows $\rho_\lambda^\text{min}$ plotted as a function of
the absolute value of the $\mu$ term (top panels) and the geometric mean of the stop masses
$\sqrt{m_{\tilde{t}_1}m_{\tilde{t}_2}}$ (bottom panels) in mSUGRA (left panels) and
mAMSB (right panels) scenarios.
It is apparent from the figure that in the cases that satisfy the observed Higgs boson mass constraint
with (near-)perfect gauge coupling unification, the $|\mu|$ term is nearly in the range of one to a few hundred TeV.
On the other hand, the stop masses, which can be taken as a rough proxy of the scale of supersymmetry,
is in the range of a few TeV to PeV.
In mSUGRA (mAMSB), these quantities are slightly towards the lower (upper) end of the aforementioned ranges.

Several comments about the viability of the parameter space with precise gauge unification are in order.
First, it is important to emphasize that the majority of this parameter space is unexplored
by the LHC or lies well beyond its reach.
Moreover, in some cases, to prevent the overproduction of gravitinos in the early universe, the reheat temperature may need to be less than about $m_{3/2}/20$, ensuring sufficient Boltzmann suppression of gravitino production \cite{Moroi:1993mb}.
In addition, due to the direct detection constraints, the neutralino LSP for Higgsino DM should be an extremely pure Higgsino \cite{Kowalska:2018toh,Martin2023}.
And, due to indirect detection constraints, wino DM seems to be experimentally less viable \cite{Cohen:2013ama,Fan:2013faa}.

%%%%%%%%%%%%%%%%%%%%%%%%%%%%%%%%%%
{\bf 5.~Conclusion.}
We have investigated the implications of precise gauge coupling unification
on the supersymmetric particle masses. We considered
the minimal supersymmetric Standard Model with a common superpartner mass,
the minimal supergravity model, and the minimal anomaly-mediated supersymmetry-breaking model.
We found that the superpartner masses are typically in the range of a few TeV and (well) beyond
in order to achieve (near-)perfect gauge coupling unification and to also obtain the correct
observed Higgs boson mass.
Even after requiring the Higgs mass conform with experiment and that gauge coupling unification is (nearly) exact, we can still identify large regions of parameter space where a Higgsino or a wino can reproduce the thermal dark matter abundance. Finally, as the analysis has made clear, unless one implements numerically too precisely and too aggressively the qualitative notions of naturalness and finetuning~\cite{Wells:2018sus}, the LHC results have had essentially no impact on the viability of supersymmetric unified theories, even precision unification. This assessment could have come out differently, depending on the gauge coupling measurements and the Higgs boson mass measurement.

\medskip
%%%%%%%%%%%%%%%%%%%%%%%%%%%%%%%%%%
\begin{acknowledgments}
{\bf Acknowledgments.}
We thank Stephen P.~Martin and Aaron Pierce for helpful discussions.
This research was supported in part through computational resources and services provided by
Advanced Research Computing (ARC), a division of Information and Technology Services (ITS) at
the University of Michigan, Ann Arbor. This work is supported by the Department of Energy under
grant number DE-SC0007859.
\end{acknowledgments}
\bibliography{ref}

\begin{thebibliography}{42}%
\makeatletter
\providecommand \@ifxundefined [1]{%
 \@ifx{#1\undefined}
}%
\providecommand \@ifnum [1]{%
 \ifnum #1\expandafter \@firstoftwo
 \else \expandafter \@secondoftwo
 \fi
}%
\providecommand \@ifx [1]{%
 \ifx #1\expandafter \@firstoftwo
 \else \expandafter \@secondoftwo
 \fi
}%
\providecommand \natexlab [1]{#1}%
\providecommand \enquote  [1]{``#1''}%
\providecommand \bibnamefont  [1]{#1}%
\providecommand \bibfnamefont [1]{#1}%
\providecommand \citenamefont [1]{#1}%
\providecommand \href@noop [0]{\@secondoftwo}%
\providecommand \href [0]{\begingroup \@sanitize@url \@href}%
\providecommand \@href[1]{\@@startlink{#1}\@@href}%
\providecommand \@@href[1]{\endgroup#1\@@endlink}%
\providecommand \@sanitize@url [0]{\catcode `\\12\catcode `\$12\catcode
  `\&12\catcode `\#12\catcode `\^12\catcode `\_12\catcode `\%12\relax}%
\providecommand \@@startlink[1]{}%
\providecommand \@@endlink[0]{}%
\providecommand \url  [0]{\begingroup\@sanitize@url \@url }%
\providecommand \@url [1]{\endgroup\@href {#1}{\urlprefix }}%
\providecommand \urlprefix  [0]{URL }%
\providecommand \Eprint [0]{\href }%
\providecommand \doibase [0]{https://doi.org/}%
\providecommand \selectlanguage [0]{\@gobble}%
\providecommand \bibinfo  [0]{\@secondoftwo}%
\providecommand \bibfield  [0]{\@secondoftwo}%
\providecommand \translation [1]{[#1]}%
\providecommand \BibitemOpen [0]{}%
\providecommand \bibitemStop [0]{}%
\providecommand \bibitemNoStop [0]{.\EOS\space}%
\providecommand \EOS [0]{\spacefactor3000\relax}%
\providecommand \BibitemShut  [1]{\csname bibitem#1\endcsname}%
\let\auto@bib@innerbib\@empty
%</preamble>
\bibitem [{\citenamefont {Dimopoulos}\ and\ \citenamefont
  {Georgi}(1981)}]{Dimopoulos:1981zb}%
  \BibitemOpen
  \bibfield  {author} {\bibinfo {author} {\bibfnamefont {S.}~\bibnamefont
  {Dimopoulos}}\ and\ \bibinfo {author} {\bibfnamefont {H.}~\bibnamefont
  {Georgi}},\ }\bibfield  {title} {\bibinfo {title} {{Softly Broken
  Supersymmetry and SU(5)}},\ }\href
  {https://doi.org/10.1016/0550-3213(81)90522-8} {\bibfield  {journal}
  {\bibinfo  {journal} {Nucl. Phys. B}\ }\textbf {\bibinfo {volume} {193}},\
  \bibinfo {pages} {150} (\bibinfo {year} {1981})}\BibitemShut {NoStop}%
\bibitem [{\citenamefont {Dimopoulos}\ \emph {et~al.}(1981)\citenamefont
  {Dimopoulos}, \citenamefont {Raby},\ and\ \citenamefont
  {Wilczek}}]{Dimopoulos:1981yj}%
  \BibitemOpen
  \bibfield  {author} {\bibinfo {author} {\bibfnamefont {S.}~\bibnamefont
  {Dimopoulos}}, \bibinfo {author} {\bibfnamefont {S.}~\bibnamefont {Raby}},\
  and\ \bibinfo {author} {\bibfnamefont {F.}~\bibnamefont {Wilczek}},\
  }\bibfield  {title} {\bibinfo {title} {{Supersymmetry and the Scale of
  Unification}},\ }\href {https://doi.org/10.1103/PhysRevD.24.1681} {\bibfield
  {journal} {\bibinfo  {journal} {Phys. Rev. D}\ }\textbf {\bibinfo {volume}
  {24}},\ \bibinfo {pages} {1681} (\bibinfo {year} {1981})}\BibitemShut
  {NoStop}%
\bibitem [{\citenamefont {Goldberg}(1983)}]{Goldberg:1983nd}%
  \BibitemOpen
  \bibfield  {author} {\bibinfo {author} {\bibfnamefont {H.}~\bibnamefont
  {Goldberg}},\ }\bibfield  {title} {\bibinfo {title} {{Constraint on the
  Photino Mass from Cosmology}},\ }\href
  {https://doi.org/10.1103/PhysRevLett.50.1419} {\bibfield  {journal} {\bibinfo
   {journal} {Phys. Rev. Lett.}\ }\textbf {\bibinfo {volume} {50}},\ \bibinfo
  {pages} {1419} (\bibinfo {year} {1983})},\ \bibinfo {note} {[Erratum:
  Phys.Rev.Lett. 103, 099905 (2009)]}\BibitemShut {NoStop}%
\bibitem [{\citenamefont {Ellis}\ \emph {et~al.}(1984)\citenamefont {Ellis},
  \citenamefont {Hagelin}, \citenamefont {Nanopoulos}, \citenamefont {Olive},\
  and\ \citenamefont {Srednicki}}]{Ellis:1983ew}%
  \BibitemOpen
  \bibfield  {author} {\bibinfo {author} {\bibfnamefont {J.~R.}\ \bibnamefont
  {Ellis}}, \bibinfo {author} {\bibfnamefont {J.~S.}\ \bibnamefont {Hagelin}},
  \bibinfo {author} {\bibfnamefont {D.~V.}\ \bibnamefont {Nanopoulos}},
  \bibinfo {author} {\bibfnamefont {K.~A.}\ \bibnamefont {Olive}},\ and\
  \bibinfo {author} {\bibfnamefont {M.}~\bibnamefont {Srednicki}},\ }\bibfield
  {title} {\bibinfo {title} {{Supersymmetric Relics from the Big Bang}},\
  }\href {https://doi.org/10.1016/0550-3213(84)90461-9} {\bibfield  {journal}
  {\bibinfo  {journal} {Nucl. Phys. B}\ }\textbf {\bibinfo {volume} {238}},\
  \bibinfo {pages} {453} (\bibinfo {year} {1984})}\BibitemShut {NoStop}%
\bibitem [{\citenamefont {Wells}(2003)}]{Wells:2003tf}%
  \BibitemOpen
  \bibfield  {author} {\bibinfo {author} {\bibfnamefont {J.~D.}\ \bibnamefont
  {Wells}},\ }\bibfield  {title} {\bibinfo {title} {{Implications of
  supersymmetry breaking with a little hierarchy between gauginos and
  scalars}},\ }in\ \href@noop {} {\emph {\bibinfo {booktitle} {{11th
  International Conference on Supersymmetry and the Unification of Fundamental
  Interactions}}}}\ (\bibinfo {year} {2003})\ \Eprint
  {https://arxiv.org/abs/hep-ph/0306127} {arXiv:hep-ph/0306127} \BibitemShut
  {NoStop}%
\bibitem [{\citenamefont {Arkani-Hamed}\ \emph {et~al.}(2005)\citenamefont
  {Arkani-Hamed}, \citenamefont {Dimopoulos}, \citenamefont {Giudice},\ and\
  \citenamefont {Romanino}}]{Arkani-Hamed:2004zhs}%
  \BibitemOpen
  \bibfield  {author} {\bibinfo {author} {\bibfnamefont {N.}~\bibnamefont
  {Arkani-Hamed}}, \bibinfo {author} {\bibfnamefont {S.}~\bibnamefont
  {Dimopoulos}}, \bibinfo {author} {\bibfnamefont {G.~F.}\ \bibnamefont
  {Giudice}},\ and\ \bibinfo {author} {\bibfnamefont {A.}~\bibnamefont
  {Romanino}},\ }\bibfield  {title} {\bibinfo {title} {{Aspects of split
  supersymmetry}},\ }\href {https://doi.org/10.1016/j.nuclphysb.2004.12.026}
  {\bibfield  {journal} {\bibinfo  {journal} {Nucl. Phys. B}\ }\textbf
  {\bibinfo {volume} {709}},\ \bibinfo {pages} {3} (\bibinfo {year} {2005})},\
  \Eprint {https://arxiv.org/abs/hep-ph/0409232} {arXiv:hep-ph/0409232}
  \BibitemShut {NoStop}%
\bibitem [{\citenamefont {Wells}(2005)}]{Wells:2004di}%
  \BibitemOpen
  \bibfield  {author} {\bibinfo {author} {\bibfnamefont {J.~D.}\ \bibnamefont
  {Wells}},\ }\bibfield  {title} {\bibinfo {title} {{PeV-scale
  supersymmetry}},\ }\href {https://doi.org/10.1103/PhysRevD.71.015013}
  {\bibfield  {journal} {\bibinfo  {journal} {Phys. Rev. D}\ }\textbf {\bibinfo
  {volume} {71}},\ \bibinfo {pages} {015013} (\bibinfo {year} {2005})},\
  \Eprint {https://arxiv.org/abs/hep-ph/0411041} {arXiv:hep-ph/0411041}
  \BibitemShut {NoStop}%
\bibitem [{\citenamefont {Arvanitaki}\ \emph {et~al.}(2013)\citenamefont
  {Arvanitaki}, \citenamefont {Craig}, \citenamefont {Dimopoulos},\ and\
  \citenamefont {Villadoro}}]{Arvanitaki:2012ps}%
  \BibitemOpen
  \bibfield  {author} {\bibinfo {author} {\bibfnamefont {A.}~\bibnamefont
  {Arvanitaki}}, \bibinfo {author} {\bibfnamefont {N.}~\bibnamefont {Craig}},
  \bibinfo {author} {\bibfnamefont {S.}~\bibnamefont {Dimopoulos}},\ and\
  \bibinfo {author} {\bibfnamefont {G.}~\bibnamefont {Villadoro}},\ }\bibfield
  {title} {\bibinfo {title} {{Mini-Split}},\ }\href
  {https://doi.org/10.1007/JHEP02(2013)126} {\bibfield  {journal} {\bibinfo
  {journal} {JHEP}\ }\textbf {\bibinfo {volume} {02}},\ \bibinfo {pages}
  {126}},\ \Eprint {https://arxiv.org/abs/1210.0555} {arXiv:1210.0555 [hep-ph]}
  \BibitemShut {NoStop}%
\bibitem [{\citenamefont {Ellis}\ and\ \citenamefont
  {Wells}(2015)}]{Ellis:2015jwa}%
  \BibitemOpen
  \bibfield  {author} {\bibinfo {author} {\bibfnamefont {S.~A.~R.}\
  \bibnamefont {Ellis}}\ and\ \bibinfo {author} {\bibfnamefont {J.~D.}\
  \bibnamefont {Wells}},\ }\bibfield  {title} {\bibinfo {title} {{Visualizing
  gauge unification with high-scale thresholds}},\ }\href
  {https://doi.org/10.1103/PhysRevD.91.075016} {\bibfield  {journal} {\bibinfo
  {journal} {Phys. Rev. D}\ }\textbf {\bibinfo {volume} {91}},\ \bibinfo
  {pages} {075016} (\bibinfo {year} {2015})},\ \Eprint
  {https://arxiv.org/abs/1502.01362} {arXiv:1502.01362 [hep-ph]} \BibitemShut
  {NoStop}%
\bibitem [{\citenamefont {Ellis}\ and\ \citenamefont
  {Wells}(2017)}]{Ellis:2017erg}%
  \BibitemOpen
  \bibfield  {author} {\bibinfo {author} {\bibfnamefont {S.~A.~R.}\
  \bibnamefont {Ellis}}\ and\ \bibinfo {author} {\bibfnamefont {J.~D.}\
  \bibnamefont {Wells}},\ }\bibfield  {title} {\bibinfo {title} {{High-scale
  supersymmetry, the Higgs boson mass, and gauge unification}},\ }\href
  {https://doi.org/10.1103/PhysRevD.96.055024} {\bibfield  {journal} {\bibinfo
  {journal} {Phys. Rev. D}\ }\textbf {\bibinfo {volume} {96}},\ \bibinfo
  {pages} {055024} (\bibinfo {year} {2017})},\ \Eprint
  {https://arxiv.org/abs/1706.00013} {arXiv:1706.00013 [hep-ph]} \BibitemShut
  {NoStop}%
\bibitem [{\citenamefont {Hebecker}\ and\ \citenamefont
  {Trapletti}(2005)}]{Hebecker:2004ce}%
  \BibitemOpen
  \bibfield  {author} {\bibinfo {author} {\bibfnamefont {A.}~\bibnamefont
  {Hebecker}}\ and\ \bibinfo {author} {\bibfnamefont {M.}~\bibnamefont
  {Trapletti}},\ }\bibfield  {title} {\bibinfo {title} {{Gauge unification in
  highly anisotropic string compactifications}},\ }\href
  {https://doi.org/10.1016/j.nuclphysb.2005.02.008} {\bibfield  {journal}
  {\bibinfo  {journal} {Nucl. Phys. B}\ }\textbf {\bibinfo {volume} {713}},\
  \bibinfo {pages} {173} (\bibinfo {year} {2005})},\ \Eprint
  {https://arxiv.org/abs/hep-th/0411131} {arXiv:hep-th/0411131} \BibitemShut
  {NoStop}%
\bibitem [{\citenamefont {Raby}\ \emph {et~al.}(2010)\citenamefont {Raby},
  \citenamefont {Ratz},\ and\ \citenamefont {Schmidt-Hoberg}}]{Raby:2009sf}%
  \BibitemOpen
  \bibfield  {author} {\bibinfo {author} {\bibfnamefont {S.}~\bibnamefont
  {Raby}}, \bibinfo {author} {\bibfnamefont {M.}~\bibnamefont {Ratz}},\ and\
  \bibinfo {author} {\bibfnamefont {K.}~\bibnamefont {Schmidt-Hoberg}},\
  }\bibfield  {title} {\bibinfo {title} {{Precision gauge unification in the
  MSSM}},\ }\href {https://doi.org/10.1016/j.physletb.2010.03.060} {\bibfield
  {journal} {\bibinfo  {journal} {Phys. Lett. B}\ }\textbf {\bibinfo {volume}
  {687}},\ \bibinfo {pages} {342} (\bibinfo {year} {2010})},\ \Eprint
  {https://arxiv.org/abs/0911.4249} {arXiv:0911.4249 [hep-ph]} \BibitemShut
  {NoStop}%
\bibitem [{\citenamefont {Krippendorf}\ \emph {et~al.}(2013)\citenamefont
  {Krippendorf}, \citenamefont {Nilles}, \citenamefont {Ratz},\ and\
  \citenamefont {Winkler}}]{Krippendorf:2013dqa}%
  \BibitemOpen
  \bibfield  {author} {\bibinfo {author} {\bibfnamefont {S.}~\bibnamefont
  {Krippendorf}}, \bibinfo {author} {\bibfnamefont {H.~P.}\ \bibnamefont
  {Nilles}}, \bibinfo {author} {\bibfnamefont {M.}~\bibnamefont {Ratz}},\ and\
  \bibinfo {author} {\bibfnamefont {M.~W.}\ \bibnamefont {Winkler}},\
  }\bibfield  {title} {\bibinfo {title} {{Hidden SUSY from precision gauge
  unification}},\ }\href {https://doi.org/10.1103/PhysRevD.88.035022}
  {\bibfield  {journal} {\bibinfo  {journal} {Phys. Rev. D}\ }\textbf {\bibinfo
  {volume} {88}},\ \bibinfo {pages} {035022} (\bibinfo {year} {2013})},\
  \Eprint {https://arxiv.org/abs/1306.0574} {arXiv:1306.0574 [hep-ph]}
  \BibitemShut {NoStop}%
\bibitem [{\citenamefont {Martin}(1998)}]{Martin:1997ns}%
  \BibitemOpen
  \bibfield  {author} {\bibinfo {author} {\bibfnamefont {S.~P.}\ \bibnamefont
  {Martin}},\ }\bibfield  {title} {\bibinfo {title} {{A Supersymmetry
  primer}},\ }\href {https://doi.org/10.1142/9789812839657_0001} {\bibfield
  {journal} {\bibinfo  {journal} {Adv. Ser. Direct. High Energy Phys.}\
  }\textbf {\bibinfo {volume} {18}},\ \bibinfo {pages} {1} (\bibinfo {year}
  {1998})},\ \Eprint {https://arxiv.org/abs/hep-ph/9709356 (updated 27 January
  2016)} {arXiv:hep-ph/9709356 (updated 27 January 2016)} \BibitemShut
  {NoStop}%
\bibitem [{\citenamefont {Giudice}\ \emph {et~al.}(1998)\citenamefont
  {Giudice}, \citenamefont {Luty}, \citenamefont {Murayama},\ and\
  \citenamefont {Rattazzi}}]{Giudice:1998xp}%
  \BibitemOpen
  \bibfield  {author} {\bibinfo {author} {\bibfnamefont {G.~F.}\ \bibnamefont
  {Giudice}}, \bibinfo {author} {\bibfnamefont {M.~A.}\ \bibnamefont {Luty}},
  \bibinfo {author} {\bibfnamefont {H.}~\bibnamefont {Murayama}},\ and\
  \bibinfo {author} {\bibfnamefont {R.}~\bibnamefont {Rattazzi}},\ }\bibfield
  {title} {\bibinfo {title} {{Gaugino mass without singlets}},\ }\href
  {https://doi.org/10.1088/1126-6708/1998/12/027} {\bibfield  {journal}
  {\bibinfo  {journal} {JHEP}\ }\textbf {\bibinfo {volume} {12}},\ \bibinfo
  {pages} {027}},\ \Eprint {https://arxiv.org/abs/hep-ph/9810442}
  {arXiv:hep-ph/9810442} \BibitemShut {NoStop}%
\bibitem [{\citenamefont {Randall}\ and\ \citenamefont
  {Sundrum}(1999)}]{Randall:1998uk}%
  \BibitemOpen
  \bibfield  {author} {\bibinfo {author} {\bibfnamefont {L.}~\bibnamefont
  {Randall}}\ and\ \bibinfo {author} {\bibfnamefont {R.}~\bibnamefont
  {Sundrum}},\ }\bibfield  {title} {\bibinfo {title} {{Out of this world
  supersymmetry breaking}},\ }\href
  {https://doi.org/10.1016/S0550-3213(99)00359-4} {\bibfield  {journal}
  {\bibinfo  {journal} {Nucl. Phys. B}\ }\textbf {\bibinfo {volume} {557}},\
  \bibinfo {pages} {79} (\bibinfo {year} {1999})},\ \Eprint
  {https://arxiv.org/abs/hep-th/9810155} {arXiv:hep-th/9810155} \BibitemShut
  {NoStop}%
\bibitem [{\citenamefont {Gherghetta}\ \emph {et~al.}(1999)\citenamefont
  {Gherghetta}, \citenamefont {Giudice},\ and\ \citenamefont
  {Wells}}]{Gherghetta:1999sw}%
  \BibitemOpen
  \bibfield  {author} {\bibinfo {author} {\bibfnamefont {T.}~\bibnamefont
  {Gherghetta}}, \bibinfo {author} {\bibfnamefont {G.~F.}\ \bibnamefont
  {Giudice}},\ and\ \bibinfo {author} {\bibfnamefont {J.~D.}\ \bibnamefont
  {Wells}},\ }\bibfield  {title} {\bibinfo {title} {{Phenomenological
  consequences of supersymmetry with anomaly induced masses}},\ }\href
  {https://doi.org/10.1016/S0550-3213(99)00429-0} {\bibfield  {journal}
  {\bibinfo  {journal} {Nucl. Phys. B}\ }\textbf {\bibinfo {volume} {559}},\
  \bibinfo {pages} {27} (\bibinfo {year} {1999})},\ \Eprint
  {https://arxiv.org/abs/hep-ph/9904378} {arXiv:hep-ph/9904378} \BibitemShut
  {NoStop}%
\bibitem [{\citenamefont {Luty}\ and\ \citenamefont
  {Sundrum}(2003)}]{Luty:2001zv}%
  \BibitemOpen
  \bibfield  {author} {\bibinfo {author} {\bibfnamefont {M.}~\bibnamefont
  {Luty}}\ and\ \bibinfo {author} {\bibfnamefont {R.}~\bibnamefont {Sundrum}},\
  }\bibfield  {title} {\bibinfo {title} {{Anomaly mediated supersymmetry
  breaking in four-dimensions, naturally}},\ }\href
  {https://doi.org/10.1103/PhysRevD.67.045007} {\bibfield  {journal} {\bibinfo
  {journal} {Phys. Rev. D}\ }\textbf {\bibinfo {volume} {67}},\ \bibinfo
  {pages} {045007} (\bibinfo {year} {2003})},\ \Eprint
  {https://arxiv.org/abs/hep-th/0111231} {arXiv:hep-th/0111231} \BibitemShut
  {NoStop}%
\bibitem [{\citenamefont {Harnik}\ \emph {et~al.}(2002)\citenamefont {Harnik},
  \citenamefont {Murayama},\ and\ \citenamefont {Pierce}}]{Harnik:2002et}%
  \BibitemOpen
  \bibfield  {author} {\bibinfo {author} {\bibfnamefont {R.}~\bibnamefont
  {Harnik}}, \bibinfo {author} {\bibfnamefont {H.}~\bibnamefont {Murayama}},\
  and\ \bibinfo {author} {\bibfnamefont {A.}~\bibnamefont {Pierce}},\
  }\bibfield  {title} {\bibinfo {title} {{Purely four-dimensional viable
  anomaly mediation}},\ }\href {https://doi.org/10.1088/1126-6708/2002/08/034}
  {\bibfield  {journal} {\bibinfo  {journal} {JHEP}\ }\textbf {\bibinfo
  {volume} {08}},\ \bibinfo {pages} {034}},\ \Eprint
  {https://arxiv.org/abs/hep-ph/0204122} {arXiv:hep-ph/0204122} \BibitemShut
  {NoStop}%
\bibitem [{\citenamefont {Profumo}\ and\ \citenamefont
  {Yaguna}(2004)}]{Profumo:2004at}%
  \BibitemOpen
  \bibfield  {author} {\bibinfo {author} {\bibfnamefont {S.}~\bibnamefont
  {Profumo}}\ and\ \bibinfo {author} {\bibfnamefont {C.~E.}\ \bibnamefont
  {Yaguna}},\ }\bibfield  {title} {\bibinfo {title} {{A Statistical analysis of
  supersymmetric dark matter in the MSSM after WMAP}},\ }\href
  {https://doi.org/10.1103/PhysRevD.70.095004} {\bibfield  {journal} {\bibinfo
  {journal} {Phys. Rev. D}\ }\textbf {\bibinfo {volume} {70}},\ \bibinfo
  {pages} {095004} (\bibinfo {year} {2004})},\ \Eprint
  {https://arxiv.org/abs/hep-ph/0407036} {arXiv:hep-ph/0407036} \BibitemShut
  {NoStop}%
\bibitem [{\citenamefont {Giudice}\ and\ \citenamefont
  {Romanino}(2004)}]{Giudice:2004tc}%
  \BibitemOpen
  \bibfield  {author} {\bibinfo {author} {\bibfnamefont {G.~F.}\ \bibnamefont
  {Giudice}}\ and\ \bibinfo {author} {\bibfnamefont {A.}~\bibnamefont
  {Romanino}},\ }\bibfield  {title} {\bibinfo {title} {{Split supersymmetry}},\
  }\href {https://doi.org/10.1016/j.nuclphysb.2004.08.001} {\bibfield
  {journal} {\bibinfo  {journal} {Nucl. Phys. B}\ }\textbf {\bibinfo {volume}
  {699}},\ \bibinfo {pages} {65} (\bibinfo {year} {2004})},\ \bibinfo {note}
  {[Erratum: Nucl.Phys.B 706, 487--487 (2005)]},\ \Eprint
  {https://arxiv.org/abs/hep-ph/0406088} {arXiv:hep-ph/0406088} \BibitemShut
  {NoStop}%
\bibitem [{\citenamefont {Pierce}(2004)}]{Pierce:2004mk}%
  \BibitemOpen
  \bibfield  {author} {\bibinfo {author} {\bibfnamefont {A.}~\bibnamefont
  {Pierce}},\ }\bibfield  {title} {\bibinfo {title} {{Dark matter in the finely
  tuned minimal supersymmetric standard model}},\ }\href
  {https://doi.org/10.1103/PhysRevD.70.075006} {\bibfield  {journal} {\bibinfo
  {journal} {Phys. Rev. D}\ }\textbf {\bibinfo {volume} {70}},\ \bibinfo
  {pages} {075006} (\bibinfo {year} {2004})},\ \Eprint
  {https://arxiv.org/abs/hep-ph/0406144} {arXiv:hep-ph/0406144} \BibitemShut
  {NoStop}%
\bibitem [{\citenamefont {Arkani-Hamed}\ \emph {et~al.}(2006)\citenamefont
  {Arkani-Hamed}, \citenamefont {Delgado},\ and\ \citenamefont
  {Giudice}}]{Arkani-Hamed:2006wnf}%
  \BibitemOpen
  \bibfield  {author} {\bibinfo {author} {\bibfnamefont {N.}~\bibnamefont
  {Arkani-Hamed}}, \bibinfo {author} {\bibfnamefont {A.}~\bibnamefont
  {Delgado}},\ and\ \bibinfo {author} {\bibfnamefont {G.~F.}\ \bibnamefont
  {Giudice}},\ }\bibfield  {title} {\bibinfo {title} {{The Well-tempered
  neutralino}},\ }\href {https://doi.org/10.1016/j.nuclphysb.2006.02.010}
  {\bibfield  {journal} {\bibinfo  {journal} {Nucl. Phys. B}\ }\textbf
  {\bibinfo {volume} {741}},\ \bibinfo {pages} {108} (\bibinfo {year}
  {2006})},\ \Eprint {https://arxiv.org/abs/hep-ph/0601041}
  {arXiv:hep-ph/0601041} \BibitemShut {NoStop}%
\bibitem [{\citenamefont {Hisano}\ \emph {et~al.}(2007)\citenamefont {Hisano},
  \citenamefont {Matsumoto}, \citenamefont {Nagai}, \citenamefont {Saito},\
  and\ \citenamefont {Senami}}]{Hisano:2006nn}%
  \BibitemOpen
  \bibfield  {author} {\bibinfo {author} {\bibfnamefont {J.}~\bibnamefont
  {Hisano}}, \bibinfo {author} {\bibfnamefont {S.}~\bibnamefont {Matsumoto}},
  \bibinfo {author} {\bibfnamefont {M.}~\bibnamefont {Nagai}}, \bibinfo
  {author} {\bibfnamefont {O.}~\bibnamefont {Saito}},\ and\ \bibinfo {author}
  {\bibfnamefont {M.}~\bibnamefont {Senami}},\ }\bibfield  {title} {\bibinfo
  {title} {{Non-perturbative effect on thermal relic abundance of dark
  matter}},\ }\href {https://doi.org/10.1016/j.physletb.2007.01.012} {\bibfield
   {journal} {\bibinfo  {journal} {Phys. Lett. B}\ }\textbf {\bibinfo {volume}
  {646}},\ \bibinfo {pages} {34} (\bibinfo {year} {2007})},\ \Eprint
  {https://arxiv.org/abs/hep-ph/0610249} {arXiv:hep-ph/0610249} \BibitemShut
  {NoStop}%
\bibitem [{\citenamefont {Cirelli}\ \emph {et~al.}(2007)\citenamefont
  {Cirelli}, \citenamefont {Strumia},\ and\ \citenamefont
  {Tamburini}}]{Cirelli:2007xd}%
  \BibitemOpen
  \bibfield  {author} {\bibinfo {author} {\bibfnamefont {M.}~\bibnamefont
  {Cirelli}}, \bibinfo {author} {\bibfnamefont {A.}~\bibnamefont {Strumia}},\
  and\ \bibinfo {author} {\bibfnamefont {M.}~\bibnamefont {Tamburini}},\
  }\bibfield  {title} {\bibinfo {title} {{Cosmology and Astrophysics of Minimal
  Dark Matter}},\ }\href {https://doi.org/10.1016/j.nuclphysb.2007.07.023}
  {\bibfield  {journal} {\bibinfo  {journal} {Nucl. Phys. B}\ }\textbf
  {\bibinfo {volume} {787}},\ \bibinfo {pages} {152} (\bibinfo {year}
  {2007})},\ \Eprint {https://arxiv.org/abs/0706.4071} {arXiv:0706.4071
  [hep-ph]} \BibitemShut {NoStop}%
\bibitem [{\citenamefont {Hryczuk}\ \emph {et~al.}(2011)\citenamefont
  {Hryczuk}, \citenamefont {Iengo},\ and\ \citenamefont
  {Ullio}}]{Hryczuk:2010zi}%
  \BibitemOpen
  \bibfield  {author} {\bibinfo {author} {\bibfnamefont {A.}~\bibnamefont
  {Hryczuk}}, \bibinfo {author} {\bibfnamefont {R.}~\bibnamefont {Iengo}},\
  and\ \bibinfo {author} {\bibfnamefont {P.}~\bibnamefont {Ullio}},\ }\bibfield
   {title} {\bibinfo {title} {{Relic densities including Sommerfeld
  enhancements in the MSSM}},\ }\href {https://doi.org/10.1007/JHEP03(2011)069}
  {\bibfield  {journal} {\bibinfo  {journal} {JHEP}\ }\textbf {\bibinfo
  {volume} {03}},\ \bibinfo {pages} {069}},\ \Eprint
  {https://arxiv.org/abs/1010.2172} {arXiv:1010.2172 [hep-ph]} \BibitemShut
  {NoStop}%
\bibitem [{\citenamefont {Baer}\ \emph {et~al.}(2011)\citenamefont {Baer},
  \citenamefont {Barger},\ and\ \citenamefont {Huang}}]{Baer:2011ec}%
  \BibitemOpen
  \bibfield  {author} {\bibinfo {author} {\bibfnamefont {H.}~\bibnamefont
  {Baer}}, \bibinfo {author} {\bibfnamefont {V.}~\bibnamefont {Barger}},\ and\
  \bibinfo {author} {\bibfnamefont {P.}~\bibnamefont {Huang}},\ }\bibfield
  {title} {\bibinfo {title} {{Hidden SUSY at the LHC: the light higgsino-world
  scenario and the role of a lepton collider}},\ }\href
  {https://doi.org/10.1007/JHEP11(2011)031} {\bibfield  {journal} {\bibinfo
  {journal} {JHEP}\ }\textbf {\bibinfo {volume} {11}},\ \bibinfo {pages}
  {031}},\ \Eprint {https://arxiv.org/abs/1107.5581} {arXiv:1107.5581 [hep-ph]}
  \BibitemShut {NoStop}%
\bibitem [{\citenamefont {Cohen}\ \emph {et~al.}(2013)\citenamefont {Cohen},
  \citenamefont {Lisanti}, \citenamefont {Pierce},\ and\ \citenamefont
  {Slatyer}}]{Cohen:2013ama}%
  \BibitemOpen
  \bibfield  {author} {\bibinfo {author} {\bibfnamefont {T.}~\bibnamefont
  {Cohen}}, \bibinfo {author} {\bibfnamefont {M.}~\bibnamefont {Lisanti}},
  \bibinfo {author} {\bibfnamefont {A.}~\bibnamefont {Pierce}},\ and\ \bibinfo
  {author} {\bibfnamefont {T.~R.}\ \bibnamefont {Slatyer}},\ }\bibfield
  {title} {\bibinfo {title} {{Wino Dark Matter Under Siege}},\ }\href
  {https://doi.org/10.1088/1475-7516/2013/10/061} {\bibfield  {journal}
  {\bibinfo  {journal} {JCAP}\ }\textbf {\bibinfo {volume} {10}},\ \bibinfo
  {pages} {061}},\ \Eprint {https://arxiv.org/abs/1307.4082} {arXiv:1307.4082
  [hep-ph]} \BibitemShut {NoStop}%
\bibitem [{\citenamefont {Fan}\ and\ \citenamefont
  {Reece}(2013)}]{Fan:2013faa}%
  \BibitemOpen
  \bibfield  {author} {\bibinfo {author} {\bibfnamefont {J.}~\bibnamefont
  {Fan}}\ and\ \bibinfo {author} {\bibfnamefont {M.}~\bibnamefont {Reece}},\
  }\bibfield  {title} {\bibinfo {title} {{In Wino Veritas? Indirect Searches
  Shed Light on Neutralino Dark Matter}},\ }\href
  {https://doi.org/10.1007/JHEP10(2013)124} {\bibfield  {journal} {\bibinfo
  {journal} {JHEP}\ }\textbf {\bibinfo {volume} {10}},\ \bibinfo {pages}
  {124}},\ \Eprint {https://arxiv.org/abs/1307.4400} {arXiv:1307.4400 [hep-ph]}
  \BibitemShut {NoStop}%
\bibitem [{\citenamefont {Baer}\ \emph {et~al.}(2016)\citenamefont {Baer},
  \citenamefont {Barger},\ and\ \citenamefont {Serce}}]{Baer:2016ucr}%
  \BibitemOpen
  \bibfield  {author} {\bibinfo {author} {\bibfnamefont {H.}~\bibnamefont
  {Baer}}, \bibinfo {author} {\bibfnamefont {V.}~\bibnamefont {Barger}},\ and\
  \bibinfo {author} {\bibfnamefont {H.}~\bibnamefont {Serce}},\ }\bibfield
  {title} {\bibinfo {title} {{SUSY under siege from direct and indirect WIMP
  detection experiments}},\ }\href {https://doi.org/10.1103/PhysRevD.94.115019}
  {\bibfield  {journal} {\bibinfo  {journal} {Phys. Rev. D}\ }\textbf {\bibinfo
  {volume} {94}},\ \bibinfo {pages} {115019} (\bibinfo {year} {2016})},\
  \Eprint {https://arxiv.org/abs/1609.06735} {arXiv:1609.06735 [hep-ph]}
  \BibitemShut {NoStop}%
\bibitem [{\citenamefont {Kowalska}\ and\ \citenamefont
  {Sessolo}(2018)}]{Kowalska:2018toh}%
  \BibitemOpen
  \bibfield  {author} {\bibinfo {author} {\bibfnamefont {K.}~\bibnamefont
  {Kowalska}}\ and\ \bibinfo {author} {\bibfnamefont {E.~M.}\ \bibnamefont
  {Sessolo}},\ }\bibfield  {title} {\bibinfo {title} {{The discreet charm of
  higgsino dark matter - a pocket review}},\ }\href
  {https://doi.org/10.1155/2018/6828560} {\bibfield  {journal} {\bibinfo
  {journal} {Adv. High Energy Phys.}\ }\textbf {\bibinfo {volume} {2018}},\
  \bibinfo {pages} {6828560} (\bibinfo {year} {2018})},\ \Eprint
  {https://arxiv.org/abs/1802.04097} {arXiv:1802.04097 [hep-ph]} \BibitemShut
  {NoStop}%
\bibitem [{\citenamefont {Co}\ \emph {et~al.}(2022{\natexlab{a}})\citenamefont
  {Co}, \citenamefont {Sheff},\ and\ \citenamefont {Wells}}]{Co:2021ion}%
  \BibitemOpen
  \bibfield  {author} {\bibinfo {author} {\bibfnamefont {R.~T.}\ \bibnamefont
  {Co}}, \bibinfo {author} {\bibfnamefont {B.}~\bibnamefont {Sheff}},\ and\
  \bibinfo {author} {\bibfnamefont {J.~D.}\ \bibnamefont {Wells}},\ }\bibfield
  {title} {\bibinfo {title} {{Race to find split Higgsino dark matter}},\
  }\href {https://doi.org/10.1103/PhysRevD.105.035012} {\bibfield  {journal}
  {\bibinfo  {journal} {Phys. Rev. D}\ }\textbf {\bibinfo {volume} {105}},\
  \bibinfo {pages} {035012} (\bibinfo {year} {2022}{\natexlab{a}})},\ \Eprint
  {https://arxiv.org/abs/2105.12142} {arXiv:2105.12142 [hep-ph]} \BibitemShut
  {NoStop}%
\bibitem [{\citenamefont {Co}\ \emph {et~al.}(2022{\natexlab{b}})\citenamefont
  {Co}, \citenamefont {Pierce}, \citenamefont {Sheff},\ and\ \citenamefont
  {Wells}}]{Co:2022jsn}%
  \BibitemOpen
  \bibfield  {author} {\bibinfo {author} {\bibfnamefont {R.~T.}\ \bibnamefont
  {Co}}, \bibinfo {author} {\bibfnamefont {A.}~\bibnamefont {Pierce}}, \bibinfo
  {author} {\bibfnamefont {B.}~\bibnamefont {Sheff}},\ and\ \bibinfo {author}
  {\bibfnamefont {J.~D.}\ \bibnamefont {Wells}},\ }\bibfield  {title} {\bibinfo
  {title} {{Discovery potential for split supersymmetry with thermal dark
  matter}},\ }\href {https://doi.org/10.1103/PhysRevD.106.095001} {\bibfield
  {journal} {\bibinfo  {journal} {Phys. Rev. D}\ }\textbf {\bibinfo {volume}
  {106}},\ \bibinfo {pages} {095001} (\bibinfo {year} {2022}{\natexlab{b}})},\
  \Eprint {https://arxiv.org/abs/2206.11912} {arXiv:2206.11912 [hep-ph]}
  \BibitemShut {NoStop}%
\bibitem [{\citenamefont {Hall}(1981)}]{Hall:1980kf}%
  \BibitemOpen
  \bibfield  {author} {\bibinfo {author} {\bibfnamefont {L.~J.}\ \bibnamefont
  {Hall}},\ }\bibfield  {title} {\bibinfo {title} {{Grand Unification of
  Effective Gauge Theories}},\ }\href
  {https://doi.org/10.1016/0550-3213(81)90498-3} {\bibfield  {journal}
  {\bibinfo  {journal} {Nucl. Phys. B}\ }\textbf {\bibinfo {volume} {178}},\
  \bibinfo {pages} {75} (\bibinfo {year} {1981})}\BibitemShut {NoStop}%
\bibitem [{\citenamefont {Raby}(2017)}]{Raby:2017ucc}%
  \BibitemOpen
  \bibfield  {author} {\bibinfo {author} {\bibfnamefont {S.}~\bibnamefont
  {Raby}},\ }\href {https://doi.org/10.1007/978-3-319-55255-2} {\emph {\bibinfo
  {title} {{Supersymmetric Grand Unified Theories}: {From Quarks to Strings via
  SUSY GUTs}}}},\ Vol.\ \bibinfo {volume} {939}\ (\bibinfo  {publisher}
  {Springer},\ \bibinfo {year} {2017})\BibitemShut {NoStop}%
\bibitem [{\citenamefont {Martin}\ and\ \citenamefont
  {Vaughn}(1994)}]{Martin:1993zk}%
  \BibitemOpen
  \bibfield  {author} {\bibinfo {author} {\bibfnamefont {S.~P.}\ \bibnamefont
  {Martin}}\ and\ \bibinfo {author} {\bibfnamefont {M.~T.}\ \bibnamefont
  {Vaughn}},\ }\bibfield  {title} {\bibinfo {title} {{Two loop renormalization
  group equations for soft supersymmetry breaking couplings}},\ }\href
  {https://doi.org/10.1103/PhysRevD.50.2282} {\bibfield  {journal} {\bibinfo
  {journal} {Phys. Rev. D}\ }\textbf {\bibinfo {volume} {50}},\ \bibinfo
  {pages} {2282} (\bibinfo {year} {1994})},\ \bibinfo {note} {[Erratum:
  Phys.Rev.D 78, 039903 (2008)]},\ \Eprint
  {https://arxiv.org/abs/hep-ph/9311340} {arXiv:hep-ph/9311340} \BibitemShut
  {NoStop}%
\bibitem [{\citenamefont {Porod}\ and\ \citenamefont
  {Staub}(2012)}]{Porod:2011nf}%
  \BibitemOpen
  \bibfield  {author} {\bibinfo {author} {\bibfnamefont {W.}~\bibnamefont
  {Porod}}\ and\ \bibinfo {author} {\bibfnamefont {F.}~\bibnamefont {Staub}},\
  }\bibfield  {title} {\bibinfo {title} {{SPheno 3.1: Extensions including
  flavour, CP-phases and models beyond the MSSM}},\ }\href
  {https://doi.org/10.1016/j.cpc.2012.05.021} {\bibfield  {journal} {\bibinfo
  {journal} {Comput. Phys. Commun.}\ }\textbf {\bibinfo {volume} {183}},\
  \bibinfo {pages} {2458} (\bibinfo {year} {2012})},\ \Eprint
  {https://arxiv.org/abs/1104.1573} {arXiv:1104.1573 [hep-ph]} \BibitemShut
  {NoStop}%
\bibitem [{\citenamefont {Porod}(2003)}]{Porod:2003um}%
  \BibitemOpen
  \bibfield  {author} {\bibinfo {author} {\bibfnamefont {W.}~\bibnamefont
  {Porod}},\ }\bibfield  {title} {\bibinfo {title} {{SPheno, a program for
  calculating supersymmetric spectra, SUSY particle decays and SUSY particle
  production at e+ e- colliders}},\ }\href
  {https://doi.org/10.1016/S0010-4655(03)00222-4} {\bibfield  {journal}
  {\bibinfo  {journal} {Comput. Phys. Commun.}\ }\textbf {\bibinfo {volume}
  {153}},\ \bibinfo {pages} {275} (\bibinfo {year} {2003})},\ \Eprint
  {https://arxiv.org/abs/hep-ph/0301101} {arXiv:hep-ph/0301101} \BibitemShut
  {NoStop}%
\bibitem [{\citenamefont {Staub}\ and\ \citenamefont
  {Porod}(2017)}]{Staub:2017jnp}%
  \BibitemOpen
  \bibfield  {author} {\bibinfo {author} {\bibfnamefont {F.}~\bibnamefont
  {Staub}}\ and\ \bibinfo {author} {\bibfnamefont {W.}~\bibnamefont {Porod}},\
  }\bibfield  {title} {\bibinfo {title} {{Improved predictions for intermediate
  and heavy Supersymmetry in the MSSM and beyond}},\ }\href
  {https://doi.org/10.1140/epjc/s10052-017-4893-7} {\bibfield  {journal}
  {\bibinfo  {journal} {Eur. Phys. J. C}\ }\textbf {\bibinfo {volume} {77}},\
  \bibinfo {pages} {338} (\bibinfo {year} {2017})},\ \Eprint
  {https://arxiv.org/abs/1703.03267} {arXiv:1703.03267 [hep-ph]} \BibitemShut
  {NoStop}%
\bibitem [{\citenamefont {Moroi}\ \emph {et~al.}(1993)\citenamefont {Moroi},
  \citenamefont {Murayama},\ and\ \citenamefont {Yamaguchi}}]{Moroi:1993mb}%
  \BibitemOpen
  \bibfield  {author} {\bibinfo {author} {\bibfnamefont {T.}~\bibnamefont
  {Moroi}}, \bibinfo {author} {\bibfnamefont {H.}~\bibnamefont {Murayama}},\
  and\ \bibinfo {author} {\bibfnamefont {M.}~\bibnamefont {Yamaguchi}},\
  }\bibfield  {title} {\bibinfo {title} {{Cosmological constraints on the light
  stable gravitino}},\ }\href {https://doi.org/10.1016/0370-2693(93)91434-O}
  {\bibfield  {journal} {\bibinfo  {journal} {Phys. Lett. B}\ }\textbf
  {\bibinfo {volume} {303}},\ \bibinfo {pages} {289} (\bibinfo {year}
  {1993})}\BibitemShut {NoStop}%
\bibitem [{\citenamefont {Martin}(2023)}]{Martin2023}%
  \BibitemOpen
  \bibfield  {author} {\bibinfo {author} {\bibfnamefont {S.~P.}\ \bibnamefont
  {Martin}},\ }\href
  {https://indico.cern.ch/event/1214022/contributions/5461067/attachments/2688809/4665483/SUSY2023_martin.pdf}
  {\bibinfo {title} {Status and future of supersymmetry}},\ \bibinfo
  {howpublished} {Presentation at the 30th International Conference on
  Supersymmetry and Unification of Fundamental Interactions} (\bibinfo {year}
  {2023})\BibitemShut {NoStop}%
\bibitem [{\citenamefont {Wells}(2019)}]{Wells:2018sus}%
  \BibitemOpen
  \bibfield  {author} {\bibinfo {author} {\bibfnamefont {J.~D.}\ \bibnamefont
  {Wells}},\ }\bibfield  {title} {\bibinfo {title} {{Naturalness,
  Extra-Empirical Theory Assessments, and the Implications of Skepticism}},\
  }\href {https://doi.org/10.1007/s10701-018-0220-x} {\bibfield  {journal}
  {\bibinfo  {journal} {Found. Phys.}\ }\textbf {\bibinfo {volume} {49}},\
  \bibinfo {pages} {991} (\bibinfo {year} {2019})},\ \Eprint
  {https://arxiv.org/abs/1806.07289} {arXiv:1806.07289 [physics.hist-ph]}
  \BibitemShut {NoStop}%
\end{thebibliography}%

\end{document}